\documentclass[letterpaper]{article} 
\usepackage{aaai24}  
\usepackage{times}  
\usepackage{helvet}  
\usepackage{courier}  
\usepackage[hyphens]{url}  
\usepackage{graphicx} 
\usepackage{natbib}  
\usepackage{caption} 
\usepackage[linesnumbered, ruled]{algorithm2e}
\usepackage{booktabs}
\usepackage{pifont}
\usepackage{multirow}
\usepackage{amsmath, amsfonts, amssymb}
\usepackage{newfloat}
\usepackage{listings}
\DeclareCaptionStyle{ruled}{labelfont=normalfont,labelsep=colon,strut=off} 
\title{Social Physics Informed Diffusion Model for Crowd Simulation}
\author{
    Hongyi Chen\textsuperscript{\rm 1,3},
    Jingtao Ding\textsuperscript{\rm 2,}\thanks{The corresponding author.},
    Yong Li\textsuperscript{\rm 2},
    Yue Wang\textsuperscript{\rm 2},
    Xiao-Ping Zhang$^{1,*}$
}
\affiliations{
    \textsuperscript{\rm 1}Shenzhen International Graduate School, Tsinghua University, Shenzhen 518055, China\\
    \textsuperscript{\rm 2}Department of Electronic Engineering, Tsinghua University, China\\
    \textsuperscript{\rm 3}Department of Strategic and Advanced Interdisciplinary Research, Peng Cheng Laboratory, Shenzhen, China\\
    {chenhy23@mails.tsinghua.edu.cn, dingjt15@tsinghua.org.cn}
}

\begin{document}

\maketitle

\begin{abstract}
Crowd simulation holds crucial applications in various domains, such as urban planning, architectural design, and traffic arrangement. 
In recent years, physics-informed machine learning methods have achieved state-of-the-art performance in crowd simulation but fail to model the heterogeneity and multi-modality of human movement comprehensively.
In this paper, we propose a social physics-informed diffusion model named SPDiff to mitigate the above gap. 
SPDiff takes both the interactive and historical information of crowds in the current timeframe to reverse the diffusion process, thereby generating the distribution of pedestrian movement in the subsequent timeframe.
Inspired by the well-known social physics model, i.e., Social Force, regarding crowd dynamics, we design a crowd interaction module to guide the denoising process and further enhance this module with the equivariant properties of crowd interactions. To mitigate error accumulation in long-term simulations, we propose a multi-frame rollout training algorithm for diffusion modeling. Experiments conducted on two real-world datasets demonstrate the superior performance of SPDiff in terms of macroscopic and microscopic evaluation metrics.
{Code and appendix are available at \url{https://github.com/tsinghua-fib-lab/SPDiff}}.
\end{abstract}

\section{Introduction}

Crowd simulation is a process of simulating the movements of a large number of people in specific scenarios, with a focus on interaction dynamics~\cite{rasouli2021pedestrian}. This technique finds its primary applications in fields such as urban planning, architectural design, and traffic management. For example, simulating how crowds move in a building under different scenarios~(i.e., crowd density, flux, etc.) enables decision-makers to assess and optimize architectural design accordingly to improve emergency response and evacuation strategies~\cite{yang2020review}. 

However, the spatio-temporal crowd trajectories are complex and heterogeneous as human behaviors are often affected by individual preferences and the surrounding environment. For example, in a shopping mall, individuals move at different speeds and follow distinct paths based on their personal interests and the mall's layout, resulting in diverse and intricate movement patterns over time. Early approaches~\cite{helbing1995social, van2011reciprocal, sarmady2010simulating, henderson1971statistics} attempted to adopt physical rule-based models to explain the underlying mechanisms behind the pedestrian movement, falling within the research domain of social physics~\cite{jusup2022social}. One notable pioneer is the Social Force Model~(SFM)~\cite{helbing1995social}, which draws inspiration from physics principles and represents pedestrians as particles influenced by various forces. With advanced deep learning, methods inspired by physics-informed machine learning~\cite{karniadakis2021physics} have achieved state-of-the-art fidelity in crowd simulation.
Examples include the PCS~(Physics-informed Crowd Simulator)~\cite{zhang2022physics} approach that replaces the core terms of SFM with GNNs~(Graph Neural Networks), and the NSP~(Neural Social Physics)~\cite{yue2022human} model that designs a learnable SFM with key parameters characterized by LSTM~(Long Short-Term Memory) based modules. 

On the other hand, the inherent uncertainty of human behavior gives rise to the indeterminacy of pedestrian trajectories, commonly referred to as the multi-modality of human movement~\cite{korbmacher2022review}. Early works~\cite{alahi2016social,mohamed2020social} made simplistic assumptions, such as Gaussian distributions, to model this multi-modality. Follow-up approaches utilized generative models such as Generative Adversarial Networks~(GANs)~\cite{gupta2018social, dendorfer2021mg, kosaraju2019social, sadeghian2019sophie} and Variational Autoencoders~(VAEs)~\cite{yue2022human, mangalam2020pecnet, ivanovic2019trajectron, chen2021personalized} to generate multimodal samples. 
In recent years, diffusion probabilistic models~\cite{NEURIPS2020_4c5bcfec} have demonstrated state-of-the-art performance in various generative tasks. This approach designs a multi-step Markov chain to reconstruct the original data distribution and generates data by stepwise denoising the noisy samples along this chain, achieving outstanding performance in capturing multimodal distributions. However, when it comes to crowd simulation, current diffusion model-based solutions~\cite{gu2022stochastic,mao2023leapfrog} are purely data-driven and thus lack guidance from prior knowledge of human movement.

Different from them, this paper comprehensively considers the two core aspects of crowd simulation and aims to design a social physics-informed diffusion model, which has two main challenges. First, \textbf{how to infuse physical knowledge regarding human movement into the diffusion model?} 
Different from diffusion models that gradually reconstruct the observed data distribution from a simple noise distribution, SFM formulates crowd movements as a many-particle dynamical system, and physical constraints are directly imposed on the observed data of every pedestrian in each timeframe. Therefore, it is difficult to infuse this knowledge into the intermediate noisy data along the diffusion process.
In contrast, current physics-guided diffusion models~\cite{xu2022geodiff, hoogeboom2022equivariant} focus on devising an equivariant diffusion framework to ensure that the generated data satisfies the corresponding geometric equivariance properties, which is distinct from the social physics knowledge~(i.e., SFM) that serves as driving force of a dynamical system. Second, \textbf{how to achieve physically consistent long-term crowd simulations with the diffusion model?} Crowd simulation is a task that involves the generation of data for multiple pedestrians and across multiple timeframes. Existing works generally adopt the one-shot generation approach of the entire sequence based on diffusion models~\cite{gu2022stochastic, tevet2022human}. However, one-shot generation cannot effectively incorporate guidance from SFM at each timeframe for each pedestrian. Moreover, it can encounter both efficiency and efficacy problems due to the high-dimensional nature of the generated data. Therefore, achieving long-term simulation and maintaining physical consistency is challenging for existing diffusion modeling frameworks.

To solve the above two challenges, we propose a conditional denoising diffusion model for crowd simulation named SPDiff that 1) includes a crowd interaction module that draws insights from the SFM to guide the denoising process, and 2) integrates strong indictive biases of equivariance properties derived from the many-particle dynamical system to enhance the model's generalization ability over transformations, leading to better performance.
Based on two designs, we further develop a multi-frame rollout training algorithm that allows the diffusion model to simulate trajectories over a defined time window and calculate the accumulated errors for updating model parameters. The resulting learning process penalizes the model for being myopic and overlooking physical consistency in the long term. 
Experiments on two real-world datasets demonstrate the significant performance improvement of SPDiff over state-of-the-art baselines, up to 18.9-37.2\% on a more difficult dataset in terms of both microscopic and macroscopic simulation realism metrics.
Further ablation studies validate SPDiff's generalization ability brought by our designed social physics-informed diffusion framework.

\begin{table*}[t]

\centering
\begin{tabular}{cccccc}
\hline
{Models} &
  {PI\textsuperscript{\rm 1}} &
  {Guidance} &
  {Indeterminacy} &
  {Approach} &
  {Optimization} \\ \hline
\multicolumn{1}{c}{STGCNN~\cite{mohamed2020social}} & \ding{55}  & -  & \ding{51} & GN\textsuperscript{\rm 2}  & End-to-End \\
\multicolumn{1}{c}{PECNet~\cite{mangalam2020pecnet}} & \ding{55}  & -  & \ding{51} & VAE            & End-to-End \\
\multicolumn{1}{c}{MID~\cite{gu2022stochastic}}    & \ding{55}  & -  & \ding{51} & DM\textsuperscript{\rm 3} & End-to-End \\
\multicolumn{1}{c}{PCS~\cite{zhang2022physics}}   & \ding{51} & SFM & \ding{55}  & -               & SFM Pretrained \\
\multicolumn{1}{c}{NSP~\cite{yue2022human}}    & \ding{51} & SFM & \ding{51} & CVAE            & Multi-staged  \\
\multicolumn{1}{c}{{SPDiff~(proposed)}} &
  \ding{51} &
  SFM, Equivariance &
  \ding{51} &
  DM\textsuperscript{\rm 3} &
  End-to-End \\ \hline
\end{tabular}
\textsuperscript{\rm 1}Physics-informed~~~~
\textsuperscript{\rm 2}Gaussian Noise~~~~
\textsuperscript{\rm 3}Diffusion Model

\caption{Comparison of deep learning based models for crowd simulation.}
\label{tab:method_compare}
\end{table*}


\section{Related Works}
\label{sec:relatedworks} 
\noindent \textbf{Crowd simulation.} 
In crowd simulation, two broad categories of methods have been identified: physics-based and data-driven methods~\cite{korbmacher2022review}. Early research focuses primarily on physics-based methods that utilize empirical social physics rules and equations to model crowd movements. The Social Force Model~(SFM)~\cite{helbing1995social} exemplifies an approach with good generalizability, representing crowd motion as a many-particle dynamical system where various forces influence pedestrians. Nevertheless, physics-based methods struggle to accurately capture micro pedestrian motion due to the complexity and indeterminacy of human behaviors, as proven in the experiments in \cite{zhang2022physics}. With the development of data science and deep learning in recent years, data-driven crowd simulation methods have been proposed to fit the distribution of microscopic human motions. 
For example, STGCNN~\cite{mohamed2020social} and PECNet~\cite{mangalam2020pecnet} utilize GNN and VAE, respectively, to predict the future trajectory distribution of pedestrians. 
However, many data-driven methods have limitations regarding generalizability to different scenarios~\cite{zhang2022physics}. Recently, physics-informed crowd simulation methods such as PCS~\cite{zhang2022physics} and NSP~\cite{yue2022human} have achieved state-of-the-art performance.
Inspired by them, we propose a novel social physics-informed diffusion model that combines the strength of generalizability in physics-based models and the distribution modeling capabilities in generative models.
We briefly summarize the main differences between SPDiff and existing works in Table~\ref{tab:method_compare}.

\noindent \textbf{Diffusion models.} The Denoising Diffusion Probabilistic Model~(DDPM)~\cite{NEURIPS2020_4c5bcfec}, standing as a prominent work in the realm of diffusion model, has gained widespread usage in the field of generation in recent years. Inspired by concepts from nonequilibrium thermodynamics~\cite{sohldickstein2015deep}, the model adds noise to original data with a certain distribution through a diffusion process modeled as a Markov chain. A neural network model is then trained to reverse the process and denoise the data, restoring the distribution of the original data from the initial noise during sampling. This kind of model has demonstrated exceptional performance in areas such as image generation~\cite{NEURIPS2020_4c5bcfec, pmlr-v139-nichol21a, NEURIPS2021_49ad23d1}, point cloud generation~\cite{luo2021diffusion}, human motion generation~\cite{tevet2022human} and spatio-temporal data generation~\cite{yuan2023spatio,zhou2023towards}.
As a representative work in trajectory prediction, MID~\cite{gu2022stochastic} models human behavior indeterminacy but does not capture pedestrian interactions. Unlike MID, our approach incorporates knowledge of social physics, considering real-time interactions and historical information as conditions. We also design a conditional diffusion framework to perform long-term simulations with multi-modality. Additionally, we have developed specific methods for multi-frame rollout training in our diffusion framework.

\noindent \textbf{Equivariant networks.} 
Problems like multi-body systems and 3D molecular structures exhibit translation and rotation symmetries. By infusing symmetry knowledge into deep learning models, the resulting equivariant networks can have much higher training efficiency~\cite{satorras2021n, Deng_2021_ICCV, pmlr-v119-kohler20a, Worrall_2017_CVPR}. 
For example, EGNN~\cite{satorras2021n} proposes an equivariant GNN network architecture that does not require computationally expensive higher-order representations for predicting a graph's state information. 
GeoDiff~\cite{xu2022geodiff} proposes a molecular conformation generation model based on the diffusion model, which possesses a rotation-translation equivariance property. 
Besides incorporating physics guidance from SFM, we further introduce equivariant design into the designed approach, considering the symmetry exhibited by the crowd motion that can be regarded as a many-particle dynamical system.

\section{Preliminaries}
\subsection{Problem Formulation}
For a group of $N$ pedestrians, crowd simulation requires consideration of the state of the crowd $Q_t=\{p_t,v_t,a_t,d,h_t\}$ at the current timeframe $t$, which comprises the positions $p_t\in\mathbb{R}^{N\times2}$, velocities $v_t\in\mathbb{R}^{N\times2}$, accelerations $a_t\in\mathbb{R}^{N\times2}$, destinations $d\in\mathbb{R}^{N\times2}$, recent historical trajectories $h_t=(p_{t-m:t}, v_{t-m:t}, a_{t-m:t})\in\mathbb{R}^{m\times N\times6}$($m$ denotes the history window length), and the positions of $M$ static obstacles in the environment $E\in\mathbb{R}^{M\times2}$. The model ${F}_\theta$ initializes from the initial state and generates the next moment's state by entering the current state, i.e.,
\begin{equation}
Q_{t+1}=F_\theta\left(Q_t,E\right),
\end{equation}
which is continuously iterated until all individuals in the crowd reach their respective destinations, completing the simulation process. 
To generate physically consistent results, we make our model directly output the acceleration $a_{t+1}\in \mathbb R^{N\times2}$ at timeframe $t+1$, which is then used to update the crowd state (position $p$ and velocity $v$) using $v_{t+1}=v_{t}+a_{t}\cdot\Delta t$ and $
p_{t+1}=p_{t}+v_{t}\cdot\Delta t$.

\subsection{Social Force Model}
From an individual perspective, the design of the dynamic mechanisms guiding pedestrian movement in our model includes destination attraction, pedestrian-pedestrian interaction, and pedestrian-obstacle interaction demonstrated in Social Force Model~\cite{helbing1995social}. Particularly, the acceleration of individual $i$ is modeled as a combination of different kinds of forces, formulated as follows,
\begin{equation}
    m_ia_i=f_{i,dest}+\sum\nolimits_{j\neq i,j\in P}f_{ji,ped}+\sum\nolimits_{o\in O}f_{oi,env}
    \label{socialforce}
\end{equation}
where $P$ and $O$ denote the set of pedestrians and the set of environmental obstacles, respectively. $f_{i,dest}$, $f_{ij,ped}$, and $f_{ik,env}$ represent the traction force from the destination to pedestrian $i$, the repulsive force from pedestrian $j$ to pedestrian $i$, and the repulsive force from obstacle $k$ to pedestrian $i$, respectively. The formula for the attractive force is given as $f_{i,dest}=m_i\frac{v_{id}n_{iD}-v_i}{\tau}$, where $v_i$ is the current velocity, $v_{id}$ is the desired walking speed, and $n_{iD}$ is the direction towards the destination. $m_i$ is a coefficient for individuals while $\tau$ is a global coefficient.

\subsection{Equivariance and Invariance} 
We say a model $\phi:X \to Y$ equivariant to transformation group $g\in G$ when:
\begin{equation}
    \phi(T_g(x)) = S_g(\phi(x)),
\end{equation}
where $T_g$ and $S_g$ are transformations on 2-D vector spaces $X$ and $Y$ for the abstract group $g$. Particularly, $\phi(T_g(x)) = \phi(x)$ stands for the invariant property of the function. In our problem, we consider the translation and rotation transformations on positions of pedestrians and obstacles, which will only lead to rotation transformations on velocities and accelerations. One of the embedding modules of our model is designed to satisfy the above equivariant constraints of positions, velocities, and accelerations on corresponding transformations.

\section{SPDiff: the Proposed Method}
\subsection{Physics Guided Conditional Diffusion Process}
\label{physicsguided}
\noindent \textbf{Framework.} 
In crowd simulation, the destinations of pedestrians are given as prior knowledge, and the destination traction force can be directly computed using known information at the current state. Based on the original SFM~\cite{helbing1995social}, our model only replaces its core terms, i.e., the repulsive forces $\sum_{j\neq i,j\in P}f_{ji,ped}+\sum_{o\in O}f_{oi,env}$, to reduce the difficulty of the stochastic prediction. We consider the neighbor pedestrians and obstacles instead of $P$ and $O$ for every pedestrian.

\begin{figure}[t]
    \centering
    \includegraphics[width=0.95\columnwidth]{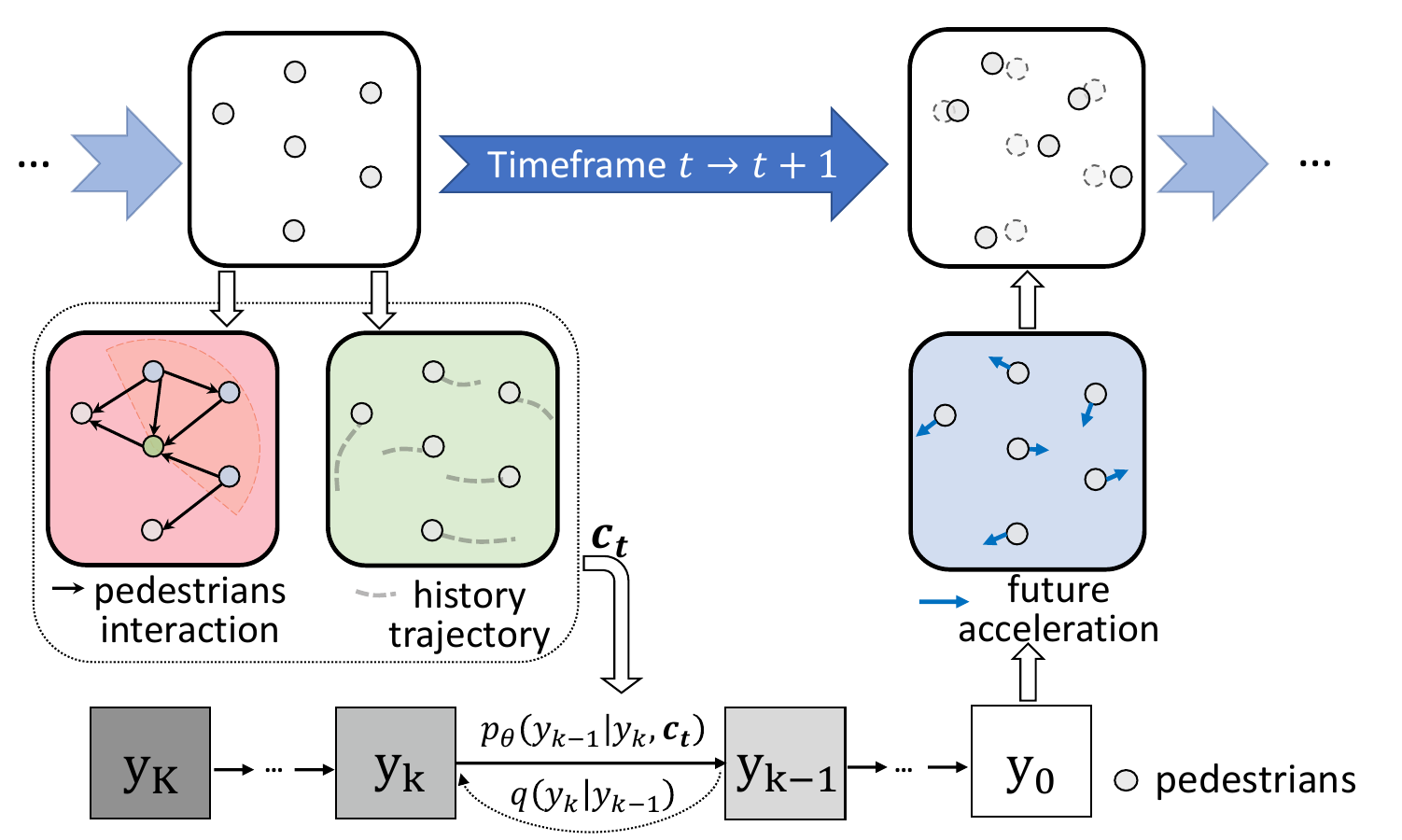}
    \caption{The overall framework of SPDiff.}
    \label{fig:framework_overall}
\end{figure}

At each time frame $t$, a graph network is employed, where interactions are formed among pedestrians based on proximity and visibility, depicted in Figure~\ref{fig:framework_overall}. Node messages in the graph represent the current states of pedestrians, including positions, velocities, and accelerations at time $t$. The proposed diffusion model utilizes the graph message and history states as conditional inputs $c_t$ and clean Gaussian noise $\mathrm{y}_K$. It predicts the future accelerations $\mathrm{y}_0=a_{t+1}$ for all existing pedestrians at the next time frame $t+1$. The pedestrians' states are then updated to simulate their progression from time $t$ to $t+1$. This iterative process continues for the entire long-term simulation. 

Due to the large number of pedestrians and extended duration, the generated data size surpasses that of human and single-pedestrian trajectory data. Coping with this substantial dataset and simulating the entire motion of large crowds pose notable challenges in model learning. Furthermore, predicting accelerations for multiple future timeframes would neglect real-time physics affecting pedestrians' movements across these frames. Consequently, unlike prevalent diffusion frameworks used for multi-timeframe data such as body motion~\cite{tevet2022human} and trajectories~\cite{gu2022stochastic}, we predict the movements for only one timeframe in each prediction step to ensure the eventual production of physics-consistent trajectories.

\noindent \textbf{Diffusion process and conditional reverse process.} 
Suppose the current timeframe is $t$. As mentioned, we predict the future acceleration $a_{t+1}$ distribution by setting it as $\mathrm{y}_0$. The forward diffusion process is defined as a Markov chain $\mathrm{y}_0,...,\mathrm{y}_k,...,\mathrm{y}_K$:
\begin{equation}
\begin{aligned}
    & q(\mathrm{y}_{1:K}|\mathrm{y}_{0})=\prod\nolimits_{k=1}^{K}  q(\mathrm{y}_k|\mathrm{y}_{k-1}), \\
    & q(\mathrm{y}_k|\mathrm{y}_{k-1})=\mathcal{N}(\mathrm{y}_k|\sqrt{1-\beta_k}\mathrm{y}_{k-1},\beta_k\mathbf{I}),
\end{aligned}
\end{equation}
where $\beta_k$ are small variance schedulers that control the noise volume added at each diffusion step $k$. So when the length of Markov chain $K$ grows, the distribution of the final variable $\mathrm{y}_K$ can be approximated to whitened isotropic Gaussian $\mathcal{N}(0,\mathbf{I})$.
The reverse process~(denoising process) is to recover the distribution of $\mathrm{y}_0$ from the pure Gaussian given the conditions $c_t$ formed by interactions and historical information. The process can be represented by the probability distribution $p(\mathrm{y}_{0:K}| c_t)=p(\mathrm{y}_K)\prod_{k=K}^{1}{p_{\theta}(\mathrm{y}_{k-1}| \mathrm{y}_k, c_t)}$, where $\mathrm{y}_K$ is the input standard Gaussian noise. Our goal is to train our reverse model $p_{\theta}(\mathrm{y}_{k-1}| \mathrm{y}_k, c_t)$ to approximate to real distribution $q(\mathrm{y}_{k-1}| \mathrm{y}_k, {\mathrm{y}}_0)$, which is tractable as it is conditioned on $\mathrm{y}_0$~\cite{NEURIPS2020_4c5bcfec}. To achieve this, we make our denoising network $\hat{\mathrm{y}}_0=f_\theta(\mathrm{y}_k,k,c_t)$ predict the desired clear sample, i.e., the acceleration $a_{t+1}$ itself, which allows us to perform our training algorithm introduced in the next section. The reverse distribution becomes:
\begin{equation}
\begin{aligned}
        p_{\theta}(\mathrm{y}_{k-1}| \mathrm{y}_k, c_t)& =q(\mathrm{y}_{k-1}| \mathrm{y}_k, \hat{\mathrm{y}}_0) \\&
    = q(\mathrm{y}_{k-1}| \mathrm{y}_k, f_\theta(\mathrm{y}_k,k,c_t))
\end{aligned}
\end{equation}

\begin{figure}[t]
    \centering
    \includegraphics[width=0.9\columnwidth]{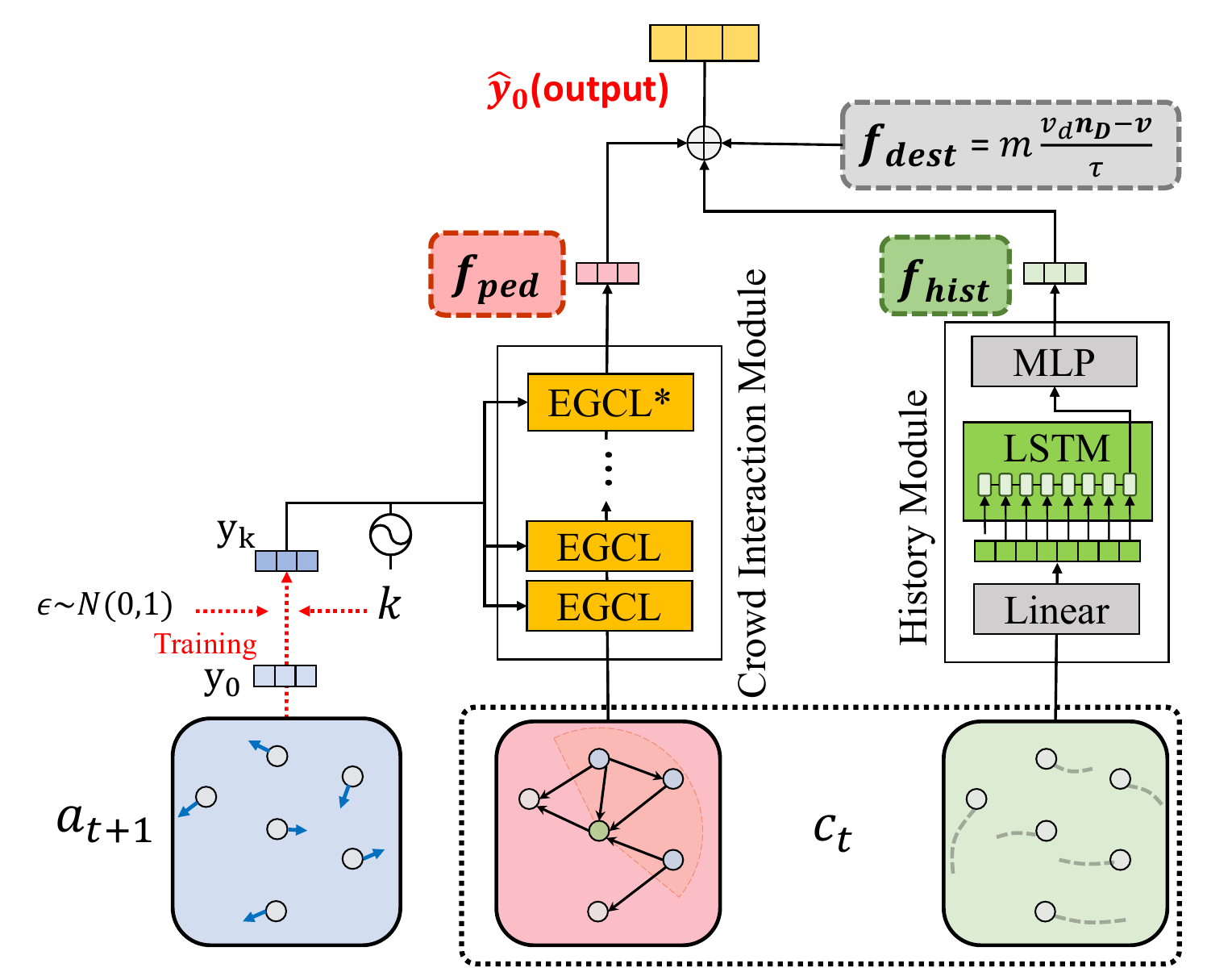}
    \caption{The detailed parameterization of the denoising \\network ($f_\theta$).}
    \label{fig:framework_detail}
\end{figure}

\subsection{Training and Sampling Algorithm}

\noindent \textbf{Multi-frame rollout training~(MRT) algorithm.}
In crowd simulation tasks, the model is required to simulate the trajectories of pedestrians at various continuous timesteps by relying solely on the initial state information. This makes it essential that the model be equipped with the ability to generalize for long-term simulation scenarios. However, training on single-step predictions is not enough due to the noise in real-world crowd data. To address this, inspired by student forcing strategy in sequence generation literatures~\cite{ranzato2015sequence}, we propose an algorithm that employs a multi-frame rollout training strategy.
When training the model, we use the model output $\hat{\mathrm{y}}_0$~(treated as $\hat{a}_{t}$ in training) of the previous timeframe $t-1$ to calculate the conditional information $\hat{c}_{t}$ for the model input at the next timeframe $t$, until it reaches a chosen training length of time $T$. We also adopt a reverse long-term discounted factor $\lambda^{T-t}$($\lambda<1$) to make the model focus on long-term accuracy~\cite{zhang2022physics}. Therefore, the corresponding loss function is calculated as follows:
\begin{equation}
\label{eq:loss}
    L = \mathbb{E}_{k,a_t}[\sum\nolimits_{t=1}^{T}\lambda^{T-t}\left|\left|a_{t}-f_\theta(\mathrm{y}_k,k,c_{t-1})\right|\right|^{2}],
\end{equation}
which follows the simple objective function demonstrated in DDPM~\cite{NEURIPS2020_4c5bcfec}. The details of the MRT in the form of pseudo-codes are provided in Appendix A in Algorithm 1.

\noindent \textbf{Sampling.}
At timeframe $t$, the diffusion model will iteratively sample from diffusion step $k=K$ to $k=1$, as the reverse process is also Markovian. In an iteration, the output of the model $f_\theta(\mathrm{y}_k,k,c_{t})$, i.e., the clean sample $\hat{\mathrm{y}}_0$, is noised back to $\mathrm{y}_{k-1}$, and the final $\mathrm{y}_0$(sampled $\hat{a}_{t+1})$ is obtained at step $k=1$, which finishes the acceleration prediction from timeframe $t$ to timeframe $t+1$. The pseudo-code is provided in Appendix A.

\subsection{Network Parameterization}
We present the model $f_\theta(\mathrm{y}_k,k,c_t)$, as illustrated in Figure~\ref{fig:framework_detail}, which is primarily composed of two modules: the crowd interaction module and the history module. 
The interaction information is processed by a module designed with equivariance property to output 2-dimensional vectors, i.e., forces, modeling the interactions of nearby pedestrians for each pedestrian in the current scene. This force vector $f_{ped}$, along with the force vector $f_{hist}$ indicating the motion effects from the history of every pedestrian, is finally added to the traction force of the destination $f_{dest}$, yields $\hat{\mathrm{y}}_0$.

    
    
\noindent \textbf{Equivariant crowd interaction module.}
We aim to obtain equivariant embedding from the interaction messages provided by the current graph. Following recent equivariant network~\cite{satorras2021n}, we propose a modified equivariant graph convolution layer~(EGCL), and our module is composed of $L$ layers of EGCL. In the $l$-th layer, node embedding $h^l$ along with corresponding position, velocity, and acceleration embeddings $p^l, v^l, a^l$ are used as inputs and are processed to update the respective embeddings $h^{l+1}, p^{l+1}, v^{l+1}, a^{l+1}$:

\begin{equation}
\label{eq:prop}
    m_{ij}=\phi_e\left(h_i^l,h_j^l,\left|\left|p_i^l-p_j^l\right|\right|^2\right),
\end{equation}
\vspace{-5px}
\begin{equation}
\label{eq:agg}
a_i^{l+1}=\phi_a\left(h_i^l\right)\mathrm{y}_{k,i}+\sum_{j\in N\left(i\right)}\frac{1}{d_{ij}}\left(p_i^l-p_j^l\right)\phi_p\left(m_{ij}\right),
\end{equation}

\begin{equation}
\label{eq:update}
v_i^{l+1}=v_i^l+a_i^{l+1},~~
p_i^{l+1}=p_i^l+v_i^{l+1},
\end{equation}
\vspace{-5px}
\begin{equation}
\label{eq:agg_update}
m_i=\sum_{j\in N\left(i\right)}m_{ij},~~
h_i^{l+1}=\phi_h\left(h_i^l,m_i\right),
\end{equation}

where $\phi$ are MLPs and $N(i)$ denotes the neighborhood of the node $i$ presented in the graph. $d_{ij}$ denotes the distance of node $i$ and $j$. $\mathrm{y}_{i,k}$ is the $i$th pedestrian's acceleration to be denoised in the noisy data input $\mathrm{y}_k$. Initial node embedding $h^0$ are the embeddings from the norms of the input velocities and accelerations($||v^{0}||$ and $||a^{0}||$), which are invariant. $p^0,v^0,a^0$ are the current positions,velocities and acceleration of the nodes(pedestrians), which are equivariant. If $h^l$ is invariant while $p^l$, $v^l$, and $a^l$ are equivariant, it can be proven that the corresponding output of the update to layer $l+1$ also satisfies the same property. Proof can be found in the appendix. In the last EGCL layer~(EGCL*), only Eq.\ref{eq:prop} and Eq.\ref{eq:agg} are used, outputting the 2-dim vector $a_i^{L}$ as the force from interactions of nearby pedestrians on pedestrian $i$. 

\noindent \textbf{History module.}
In crowd simulation, the movement of pedestrians can often be influenced by their historical trajectories. This can be attributed to the prior knowledge that humans tend to avoid changing their movement too much to conserve energy. Therefore, in each simulation iteration, we collect each pedestrian's movement states(position, velocity, acceleration) over the previous 8 frames as input $h_t\in\mathbb R^{8\times N\times 6}$. The 8-length sequence is upsampled using linear layers and then encoded using an $\mathrm{LSTM}$, which outputs the hidden embedding of the last token, decoded by an MLP, as shown in the following formula:
\begin{equation}
    f_{hist}=\mathrm{MLP}(\mathrm{LSTM}(\mathrm{Linear}(h_t))).
\end{equation}

\section{Experiments}
\subsection{Experiment Setup}
\noindent \textbf{Datasets.}
We conduct crowd simulation evaluation experiments of the model on two open-source datasets: the GC and the UCY datasets. The two datasets differ in scenarios, scale, duration, and pedestrian density, allowing us to verify the model's generalization performance. Following the approach of PCS~\cite{zhang2022physics}, we select trajectory data with rich pedestrian interactions~($> 200$ pedestrians per minute) of 300s duration from the GC dataset and 216s duration from the UCY dataset for training and testing. Please refer to Appendix B for detailed information.
\begin{table*}[]
\centering
\resizebox{\textwidth}{!}{%
\normalsize
{\fontsize{14pt}{11pt}\selectfont
\begin{tabular}{lcccccccccccc}

\toprule
 &
   &
  \multicolumn{5}{c}{\textbf{GC}} &
  \multicolumn{5}{c}{\textbf{UCY}} &
   \\ 
\multirow{-2}{*}{\textbf{Group}} &
  \multirow{-2}{*}{\textbf{Models}} &
  \textbf{MAE} &
  \textbf{OT} &
  \textbf{MMD} &
  \textbf{DTW} &
  \textbf{\#Col} &
  \textbf{MAE} &
  \textbf{OT} &
  \textbf{MMD} &
  \textbf{DTW} &
  \textbf{\#Col} &
  \multirow{-2}{*}{\#Params} \\ \midrule
 &
  CA &
  2.7080 &
  5.4990 &
  0.0620 &
  - &
  1492 &
  8.3360 &
  79.4200 &
  2.0220 &
  - &
  4504 &
   \\
\multirow{-2}{*}{Physics-based} &
  SFM &
  1.2590 &
  2.1140 &
  0.0150 &
  - &
  \textbf{622} &
  2.5390 &
  6.5710 &
  0.1290 &
  - &
  434 &
   \\ \midrule
 &
  STGCNN &
  8.1608 &
  15.8372 &
  0.5296 &
  5.1438 &
  2076 &
  7.5121 &
  18.7721 &
  0.5149 &
  5.1695 &
  1348 &
  7.6K \\
 &
  PECNet &
  2.0669 &
  4.3054 &
  0.0397 &
  0.7431 &
  1142 &
  3.9674 &
  16.1412 &
  0.1504 &
  2.0986 &
  1092 &
  2.1M \\
\multirow{-3}{*}{Data-driven} &
  MID &
  8.4257 &
  35.1797 &
  0.3737 &
  4.2773 &
  1620 &
  8.2915 &
  47.8711 &
  0.4384 &
  4.7109 &
  1076 &
  2.5M \\ \midrule
 &
  PCS &
  1.0320 &
  1.5963 &
  0.0126 &
  0.4378 &
  764 &
  {\underline {2.3134}} &
  {\underline {6.2336}} &
  {\underline {0.1070}} &
  {\underline {0.9887}} &
  \textbf{238} &
  0.6M \\
 &
  NSP &
  {\underline {0.9884}} &
  {\underline {1.4893}} &
  {\underline {0.0106}} &
  \textbf{0.3329} &
  {\underline {734}} &
  2.4006 &
  6.3795 &
  0.1199 &
  0.9965 &
  380 &
  2.5M \\
\multirow{-3}{*}{Physics-informed} &
  \textbf{Ours} &
  \textbf{0.9116} &
  \textbf{1.3925} &
  \textbf{0.0092} &
  {\underline {0.3332}} &
  810 &
  \textbf{1.8760} &
  \textbf{4.0564} &
  \textbf{0.0671} &
  \textbf{0.7541} &
  {\underline {372}} &
  0.2M \\ \bottomrule
\end{tabular}%
}
}
\textreferencemark ~~The results of CA and SFM are directly copied from \cite{zhang2022physics} without evaluating DTW.
\caption{Overall performance comparison.}
\label{tab:overall_perf}
\end{table*}

\noindent \textbf{Baseline methods.} 
We divide the baseline methods into physics-based, data-driven, and physics-informed methods. Within the physics-based methods, we choose the widely-used Social Force Model~(SFM)~\cite{helbing1995social} and Cellular Automaton(CA)~\cite{sarmady2010simulating} for comparison. Within the data-driven methods, we select three representative approaches recently published, including STGCNN~\cite{mohamed2020social} which utilizes graph convolutional neural networks to compute a spatio-temporal embedding, PECNet~\cite{mangalam2020pecnet} which uses VAE to sample multi-modal endpoints and MID~\cite{gu2022stochastic} which is based on the diffusion framework to model indeterminacy. For physics-informed methods, we select PCS~\cite{zhang2022physics}, whose backbone is graph networks, and NSP~\cite{yue2022human}, based on sequence prediction models combined with CVAE. The details of the implementation of baselines are in Appendix B.

\noindent \textbf{Experiment settings.}
We temporally split the datasets into training and testing sets, with a training-to-testing ratio of 4:1 for the GC dataset and 3:1 for the UCY dataset. We assess the performance using four metrics. To measure the microscopic simulation accuracy compared to the ground truth, we employ the Mean Square Error~(MAE) and the Dynamic Time Warping~(DTW), which is commonly used to measure time-dependent sequences' similarity and is a reliable metric for assessing trajectory similarity in shape
. As performing quantitative validation is also essential~\cite{10.1109/TVCG.2016.2642963, 10.1145/3386569.3392407}, we test on the \#Col~(number of collisions), characterizing the simulation’s realness. At a macroscopic level, we consider the distribution aspect and selected Optimal Transport~(OT)~\cite{villani2021topics} and Maximum Mean Discrepancy~(MMD)~\cite{gretton2012kernel}, widely used in measuring the distribution similarity of simulated physical processes~\cite{sanchez2020learning}, to measure the difference between the simulated trajectory distribution and the ground truth. We also evaluate the visualization performances of our method and some baselines, which can be found in supplementary materials.
We have full details on metrics and implementations in Appendix B.

\subsection{Overall Performance}
As shown in Table~\ref{tab:overall_perf}, we present the results of SPDiff and the baseline methods on two real-world datasets. SPDiff outperforms other existing methods, showing a relative improvement of 6.5\%-13.5\% on the MAE, OT, and MMD metrics for the GC dataset. On the UCY dataset, it achieves an improvement of 18.9\%-37.2\% across all metrics. Specifically, we have the following observations.
First, our method, guided by SFM, outperforms physics-based methods, exhibiting better fitting of pedestrian movement distributions using real-world data compared to pure social physics equations. Second, our model surpasses data-driven methods by incorporating social physics. Particularly, SPDiff outperforms the diffusion-based MID thanks to our design of the training mechanism applied to the diffusion framework. Among the data-driven models proposed for trajectory prediction tasks, only PECNet shows a comparable performance due to its dedicated design for handling trajectory endpoints. Third, it is notable that physics-informed methods outperform all three categories, highlighting the significance of physics-informed approaches in crowd simulation. By successfully applying the diffusion model to crowd simulation, our method outperforms the other two on most metrics, with the only deficiency observed in \#Col.

As can be seen, most methods performed better on the GC dataset than the UCY dataset, indicating that the GC dataset is easier to fit since pedestrians in the UCY dataset have a larger variance of speed~(See Appendix). Meanwhile, our method exhibits better improvement in UCY than in GC, demonstrating the superior ability of our model to handle difficult-to-learn datasets. 

In addition, we compare the number of trainable parameters of the DL-based methods and show that our method achieves the best performance while utilizing only 8\%/33\% of parameters compared with competitive baselines NSP/PCS. This owes to the equivariant design that reduces the parameter cost of learning the rotation-equivariant interaction information.

\subsection{Rollout Error Analysis of the Simulation}
\label{sec:rollouterror}
\begin{figure}[t]
    \centering
    \includegraphics[width=.42\columnwidth]{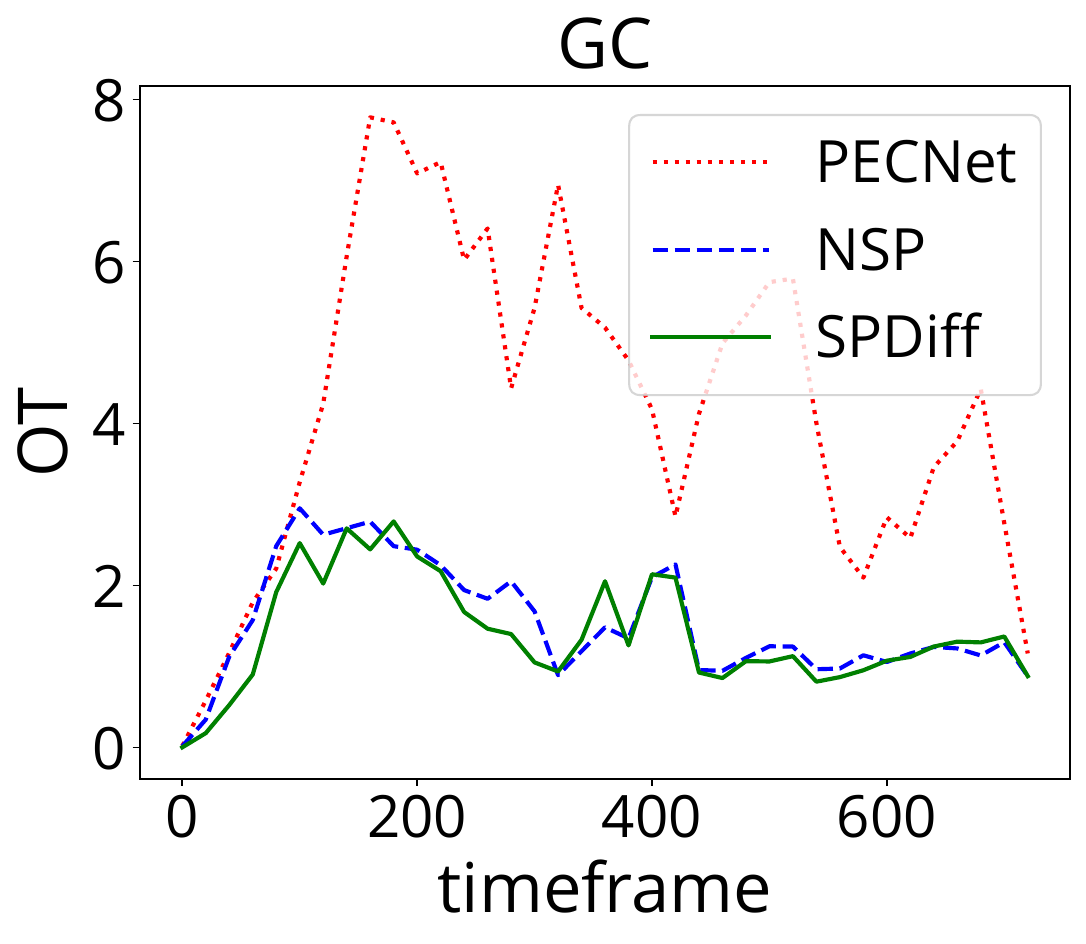}
    \includegraphics[width=.45\columnwidth]{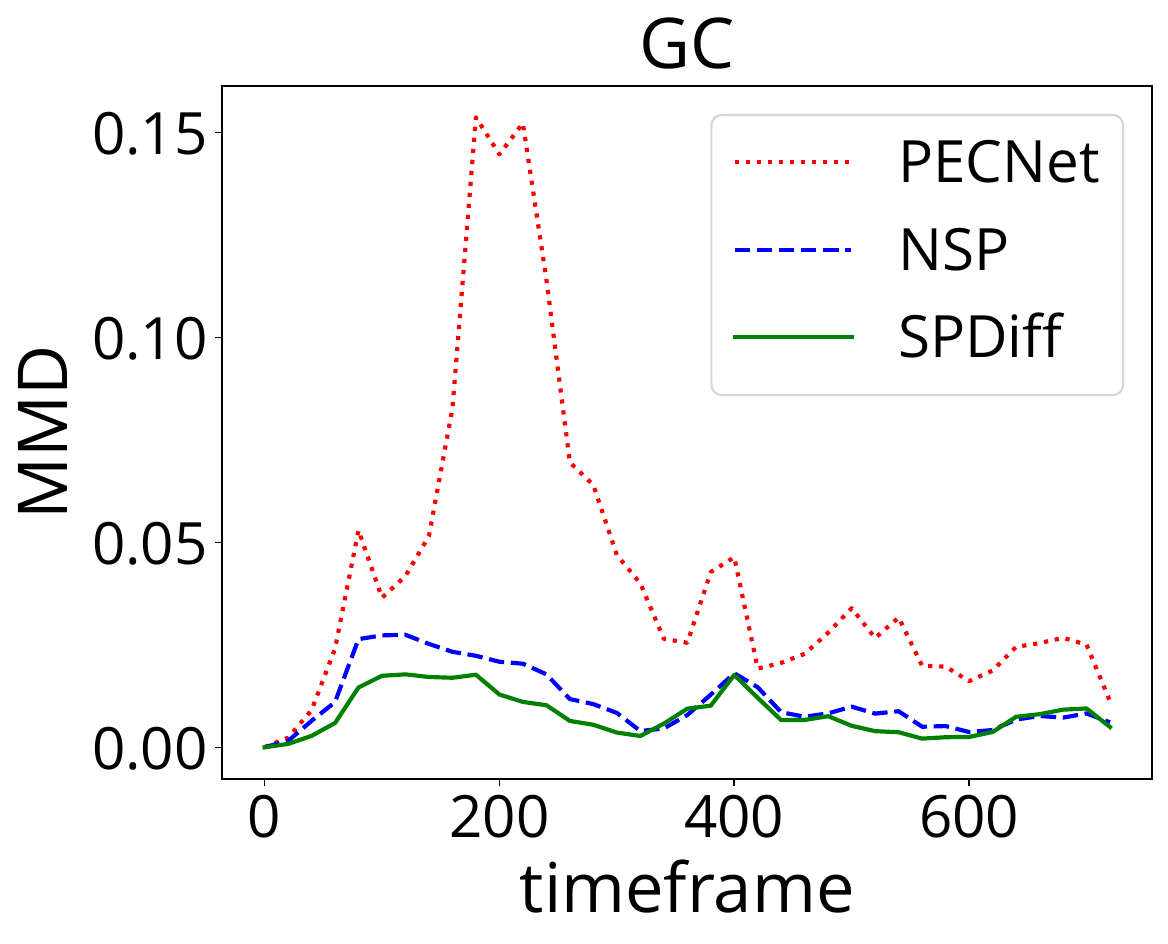}\\
    \includegraphics[width=.44\columnwidth]{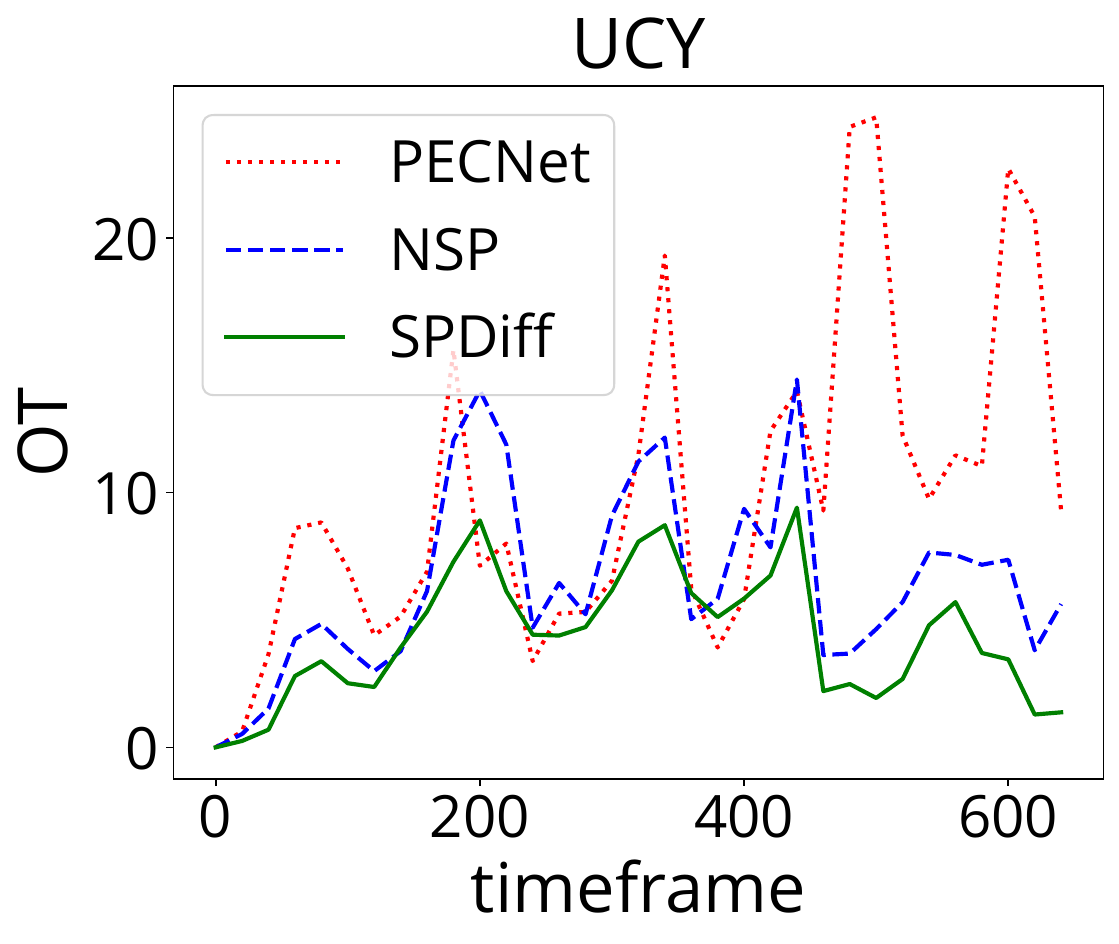}
    \includegraphics[width=.45\columnwidth]{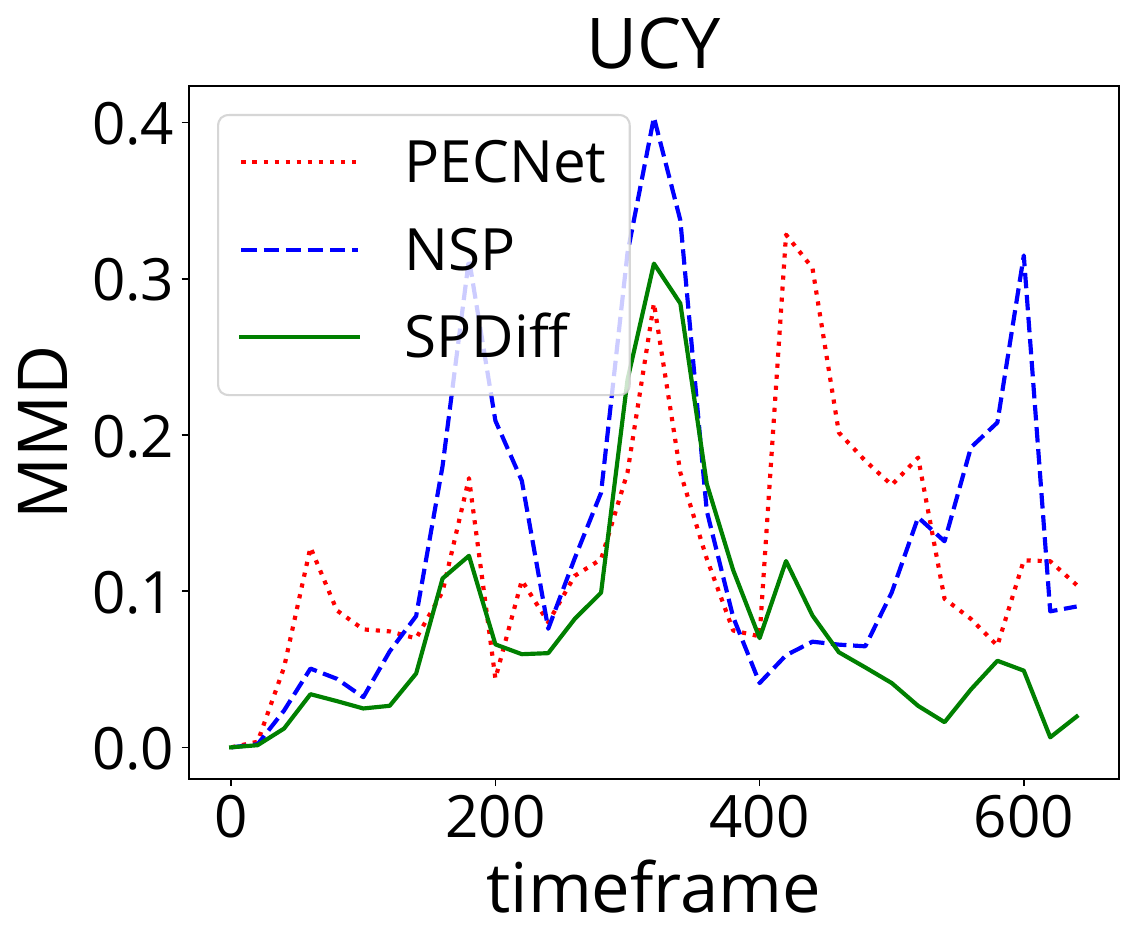}
    \caption{Rollout error as a function of frames, using OT and MMD as metrics.}
    \label{fig:rollout_metric}
\end{figure}

To further investigate simulation performance in each detailed timeframe, we examine the variations of the distributional metrics OT and MMD during the simulation rollout. Figure~\ref{fig:rollout_metric} illustrates the results on GC and UCY datasets in comparison to the baselines PECNet and NSP, which perform the best in their categories. The figures reveal oscillating trends in the metrics with alternating increases and decreases. The increases can be attributed to the cumulative error generated during the multi-frame rollout. Meanwhile, the distributional error at the pedestrian endpoints diminishes as the pedestrians are set to their real endpoints at their final appearance in the simulation.

The following observations can be gleaned: 1) Data-driven methods~(PECNet), show a higher accumulation of errors over a longer duration. In contrast, physics-informed methods~(SPDiff and NSP), which integrate constraints derived from physical knowledge, can control error accumulation within a certain range. 2) 
Our approach has lower cumulative error over time than the physics-informed NSP method, which is strongly constrained by SFM equations and relies only on historical trajectory information for modeling multi-modality. In contrast, our diffusion model, not rigidly confined by SFM representations, can learn more realistic distributions from the data and effectively model multi-modality by leveraging both historical trajectories and interaction information.

Moreover, to examine our model's generalizability beyond its training distributions, we test the performance on some scenarios picked from the SDD(Standford Drone Dataset) dataset using methods trained on the GC dataset and prove the good generalizability of SPDiff. Details and results can be found in the supplementary materials.

\begin{table*}[ht]
\centering
\begin{tabular}{ccccccccc}
\toprule
 & \multicolumn{4}{c|}{{GC}}    & \multicolumn{4}{c}{{UCY}}   \\
 & {MAE}    & {OT}     & {MMD}    & \multicolumn{1}{c|}{DTW}    & {MAE}    & {OT}     & {MMD}    & {DTW}    \\ \midrule
\textbf{Ours}   & \textbf{0.9116} & \textbf{1.3925} & \textbf{0.0092} & \multicolumn{1}{c|}{\textbf{0.3332}} & \textbf{1.8760} & \textbf{4.0564} & \textbf{0.0671} & \textbf{0.7541} \\
w/o Social Physics                    & 3.3102 & 13.6530 & 0.0637 & \multicolumn{1}{c|}{1.6517} & 3.5404 & 12.9325 & 0.1541 & 2.0016 \\
w/o History Variant                   & 1.0834 & 1.8482  & 0.0154 & \multicolumn{1}{c|}{0.3883} & 2.3340 & 6.2837  & 0.1171 & 1.1055 \\
w/o Multistep Rollout Training                               & 1.0214 & 1.6790  & 0.0141 & \multicolumn{1}{c|}{0.4032} & NC      & NC       & NC      & NC      \\ \bottomrule
\end{tabular}%
\caption{Ablation study on different parts of model design~(``NC'' denotes ``not converged'').}
\label{tab:ablation}
\end{table*}
\begin{figure}[t]
    \centering
    \includegraphics[width=.42\columnwidth]{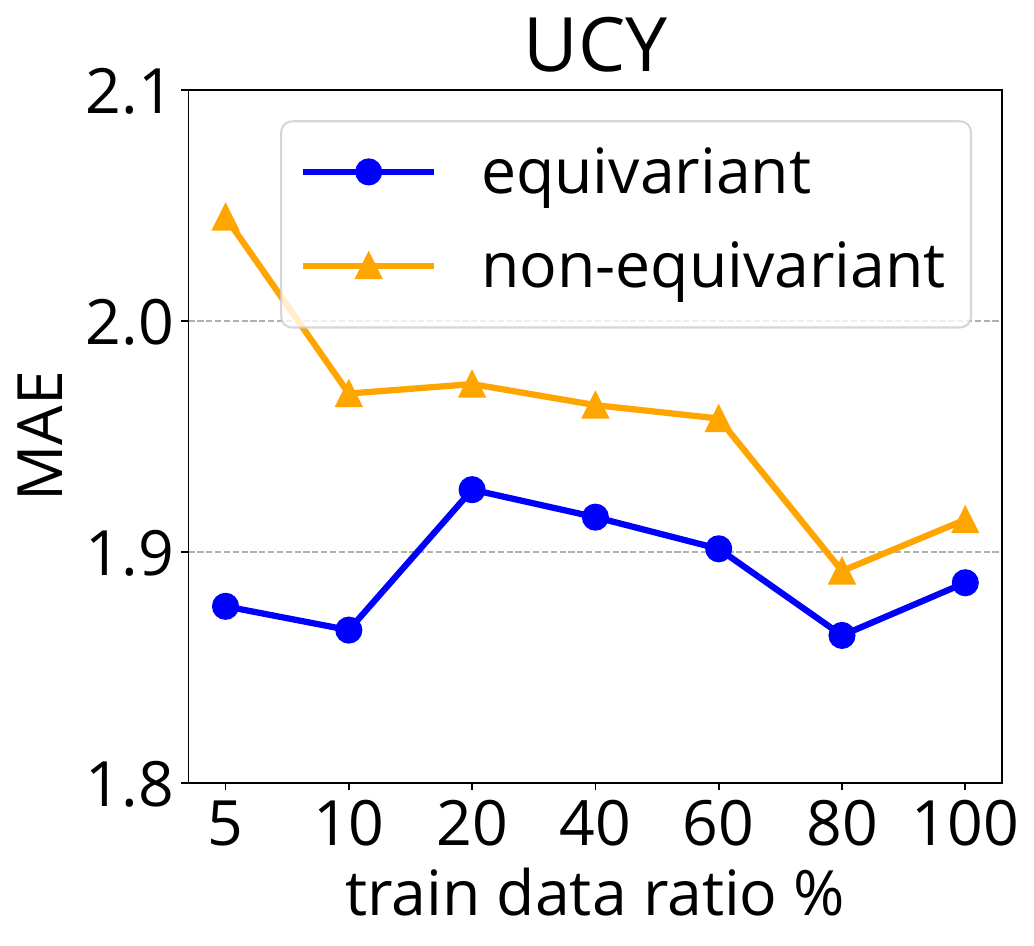}
    \includegraphics[width=.42\columnwidth]{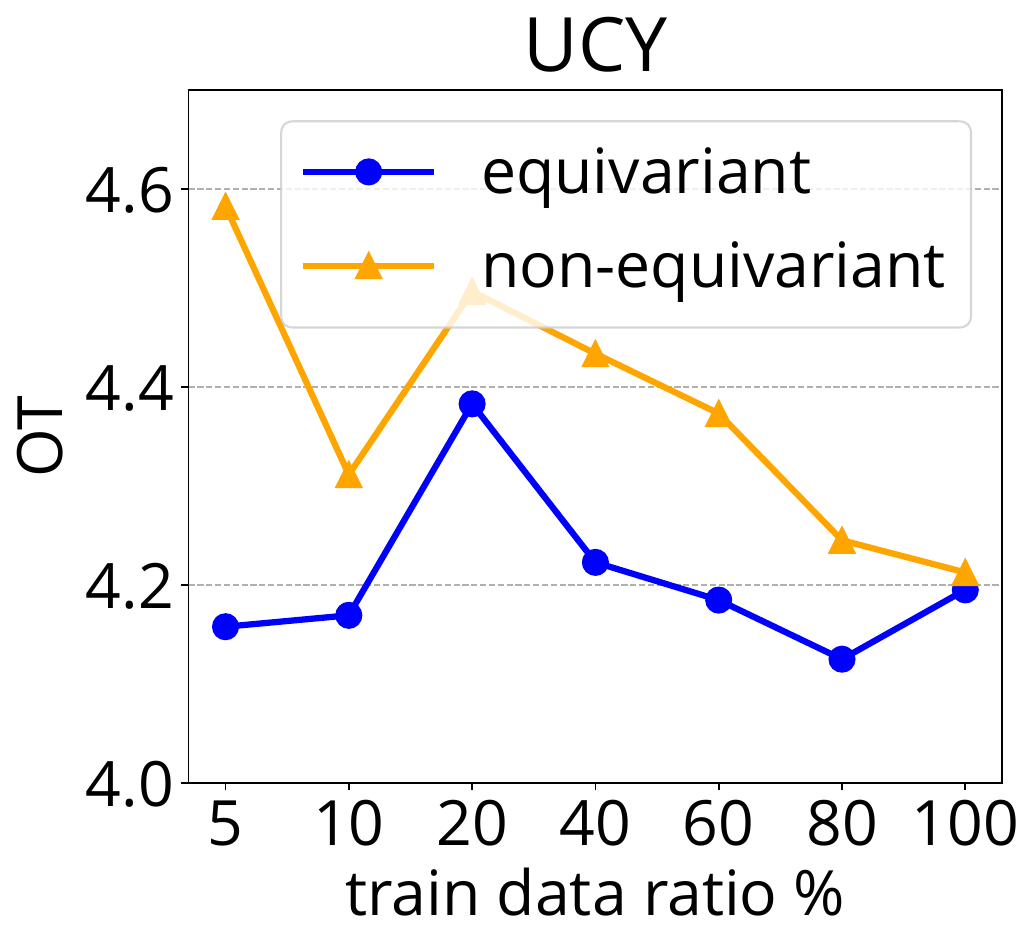}
    \includegraphics[width=.45\columnwidth]{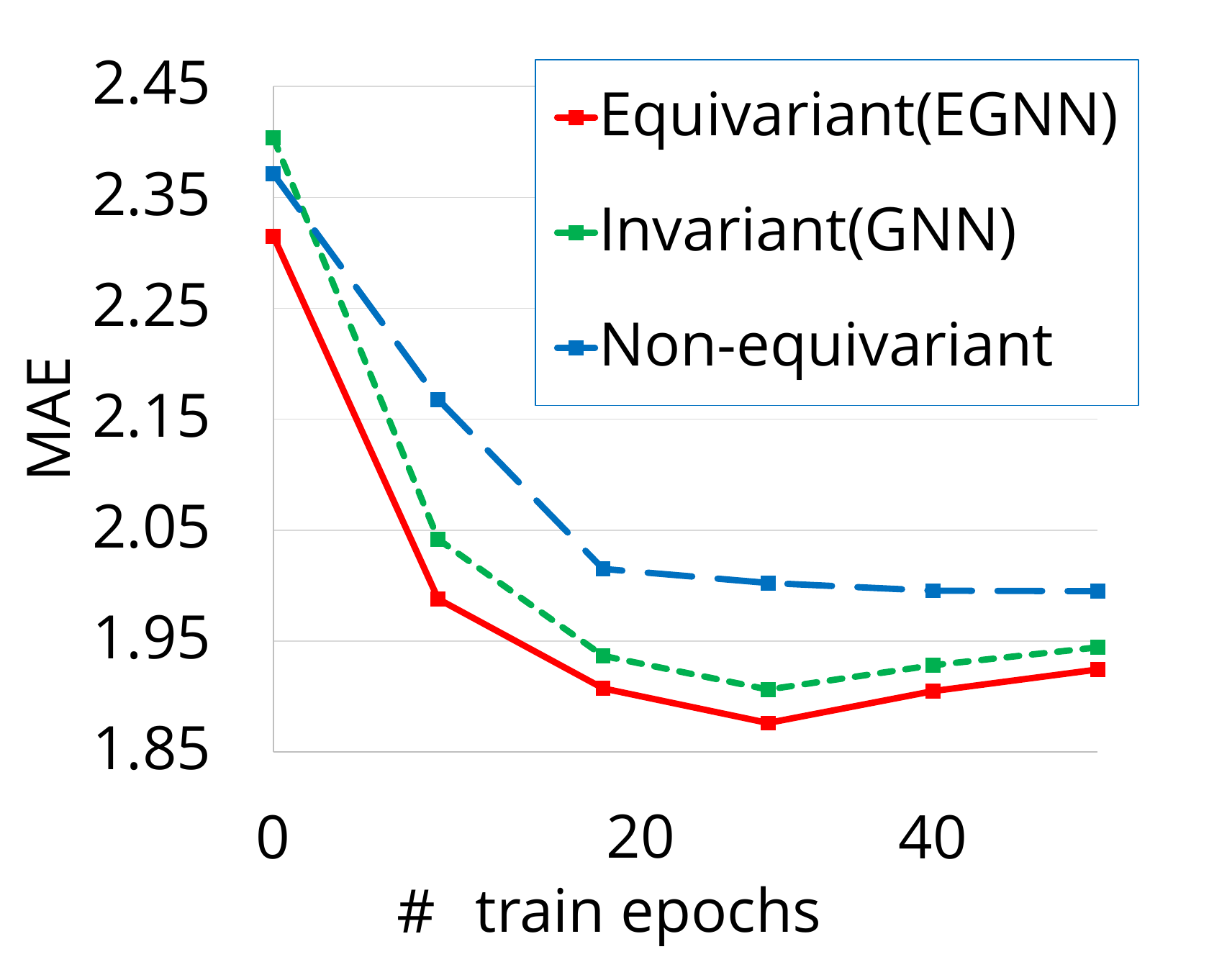}
    \includegraphics[width=.45\columnwidth]{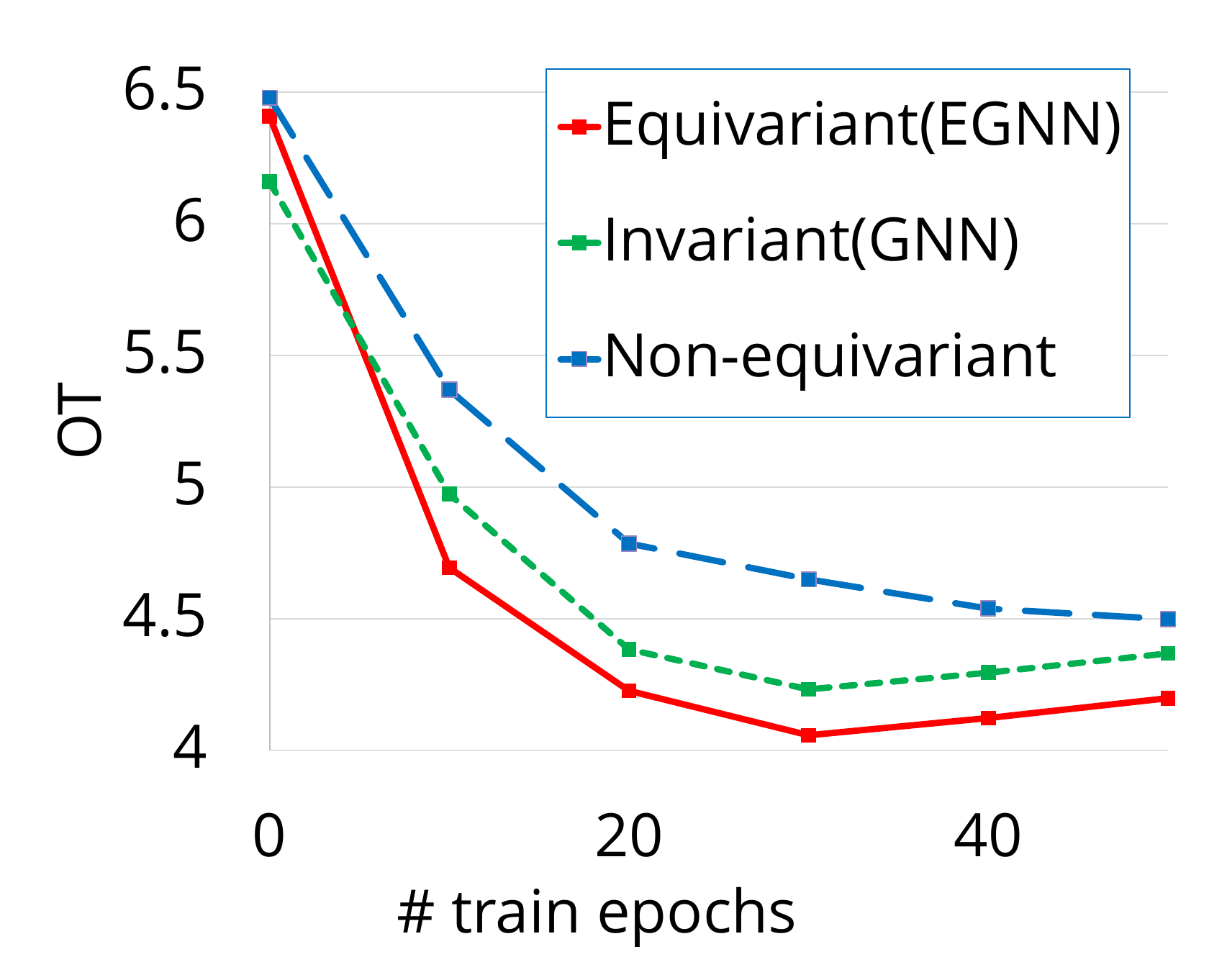}\\
    \caption{Top: Test performance under different training sample sizes on the UCY dataset. Bottom: Test performance under different training epochs on the UCY dataset.}
    \label{fig:equivariance_study}
\end{figure}

\subsection{Ablation Study}
\noindent \textbf{Effect of network modules.}
We further explore the performance contribution of each key design of our approach to investigate their necessity, and we consider four variants, as shown in Table~\ref{tab:ablation}. The \textit{w/o history variant} removes the history module and the corresponding inputs, while the \textit{w/o social physics} variant excludes modules related to social physics knowledge(crowd interaction module and social force guidance). And finally, the \textit{w/o Multistep Rollout Training} variant only utilizes a single timeframe of the model output for loss calculation and gradient descent. 

We present the performance results of the aforementioned version in Table~\ref{tab:ablation}. Note that without MRT, the metrics cannot converge on the UCY dataset. As can be seen, removing any components leads to a certain decrease in performance, demonstrating the effectiveness of each design. Most importantly, the largest performance loss is observed when removing the design related to social physics guidance, highlighting the necessity of incorporating social physics knowledge in crowd simulation. Compared with social physics, the history module is less important as human motions highly depend on the current context instead of history. Finally, in the UCY dataset, which is more challenging to fit, the metrics fail to converge without employing the MRT algorithm, demonstrating the necessity of the long-term training techniques employed in the diffusion framework.

\noindent \textbf{Effect of equivariant design.}
To investigate the impact of the inductive bias brought by the equivariant design, we conducted a performance comparison of SPDiff with two degenerations on the equivariant crowd interaction module:~1)~\textit{an invariant GNN module}, which simply replaces EGCLs with modified GCLs(Graph Convolutional Layers) encoding the relative state information to ensure invariance, and 2)~\textit{a non-equivariant crowd interaction module} inspired by that of PCS~\cite{zhang2022physics}. This module adopts a multi-layer perceptron with residual bypass~(ResMLP) to encode the relative state information between pedestrians and their neighbors. We replace the multiple EGCLs with this design in the non-equivariant crowd interaction encoder and adjust the number of parameters at a comparable level. We present their test performance under different training samples and epochs \textit{w.r.t} MAE and OT on the UCY dataset, covering microscopic and macroscopic error evaluation.

As shown at the top of Figure~\ref{fig:equivariance_study}, our method consistently outperforms the modified model with the non-equivariant interaction module under nearly all training sample ratios and remains the performance even when using 5\% of the training data. Specifically, at 5\%, SPDiff exhibits very little MAE degradation compared to the 100\% point, with a maximum decrease of only 2.5\%. Meanwhile, the equivariant design has gained at most 13.2\% of increase in MAE and 22\% of improvement in OT compared to the non-equivariant design, illustrating that our model possesses enhanced generalization ability over rotations with the help of the equivariant design. Bottom figures also show the better performance of our model compared with the invariant and non-equivariant at each converged point, with a 1.6\% of increase in MAE, a 4.1\% of increase in OT and a 13.7\% of increase in MMD compared to the second best. Moreover, it can be gleaned that models leveraging equivariance or invariance converge faster than the non-equivariant(also non-invariant) module, demonstrating the training efficiency improvement brought by our equivariant design.

\section{Conclusion}
This paper proposes a novel conditional denoising diffusion model SPDiff that can effectively leverage interaction dynamics for crowd simulation with a physics-guided diffusion process.
Motivated by the well-known SFM, our equivariant crowd interaction module and multi-frame rollout training algorithm achieve macro-and-micro realism and long-term consistency in simulation. 
Experiments on two real-world datasets demonstrate SPDiff’s superiorities in achieving the best performance with fewer parameters.

\newpage

\section{Acknowledgments}
This work was supported in part by the National Key Research and Development Program of China under 2022YFF0606904 and the
National Natural Science Foundation of China under U21B2036, U20B2060.

\bibliography{aaai24}

\end{document}